\documentclass[11pt,twoside]{article} \usepackage{asp2010}

\resetcounters

\begin{document}

\title{Numerical challenges in kinetic simulations of three-wave interactions}
\author{Urs~Ganse,$^1$ Patrick~Kilian,$^1$, Stefan~Siegel$^1$ and
Felix~Spanier$^1$ \affil{$^1$Lehrstuhl f\"ur Astronomie, Universit\"at
W\"urzburg, Emil-Fischer-Stra\ss{}e 31, 97074 W\"urzburg, Germany} }

\begin{abstract}
	Generation of radio bursts in CME foreshock regions and turbulent cascades in
	the solar wind are assumed to be results of three-wave interaction processes
	of dispersive plasma modes. Using our Particle in Cell code ACRONYM, we have
	studied the behaviour of kinetic wavemodes in the presence of beamed electron
	populations, with a focus on type II radio burst emission processes. We
	discuss the numerical challenges in generating and analyzing
	self-consistently evolving wave coupling processes with a PiC-Code and
	present preliminary results of said project.  \end{abstract}

\section{Three wave interaction}

Plasma behaviour in general is usually described by a number of nonlinear
differential equations. Depending on application, these may be the MHD
equations, the Vlasov-Maxwell system or the multifluid equations. Since
analytic general solutions to these equations are not available, a typical
solution ansatz involves linearized wave solutions which are assumed to be
small relative to the fixed background. 

In first order, this yields the canonical linear wave solutions of the
corresponding system of equations, known as ``MHD waves'' or ``kinetic waves'',
for which analytic expressions are well-documented in literature.

However if emission, absorption or interaction processes of these waves are to
be investigated, next-higher order terms of the equations need to be taken into
account, leading to three wave interaction processes of waves $A$, $B$ and $C$,
which may have the forms \begin{eqnarray} A + B \rightarrow C\\ A \rightarrow B
	+ C, \end{eqnarray} which are known as wave coalescing- and decay processes,
	respectively.
Analytic derivations of interaction rates for these processes are often quite
involved, and only available for limited subsets of certain wave families
\citep{Melrose, SpanierVainio09}.

The ability to reproduce the complete nonlinear dynamics of a plasma system,
and the possibility to analyze all physical properties of wave-like phenomena
within this system make numerical simulations extremely valuable for these
studies.

\section{Specific Problem: Emission of Type II Radio Bursts}

\begin{figure} \begin{center}
	\includegraphics[width=\textwidth]{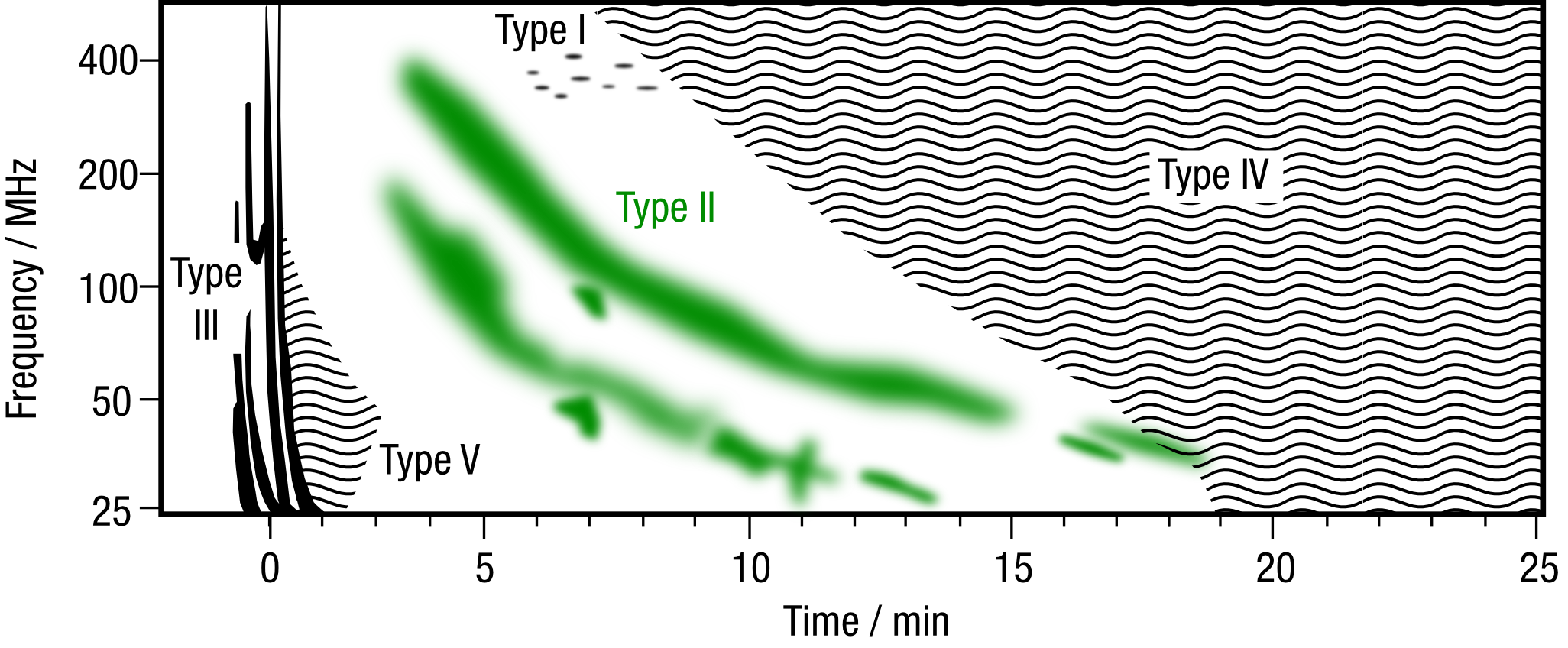} \end{center}
	\caption{Schematic classification and features of solar radio bursts. Of
	particular interest here are the type II and III radio bursts, which show
	multi-banded narrowband emission drifting towards lower frequency.}
	\label{fig:radiobursttypes} \end{figure}

One wave emission process that has been of large importance in this context is
the emission of solar radio bursts of type II and III. 

These transient radio phenomena from the sun have been observed since the
earliest days of solar radio observations in the 1950ies \citep{WildMcCready}.
They are characterized by narrowband multi-banded radio emissions, which are
rapidly drifting downwards in frequency. In the case of type III bursts, this
has been identified with energetic electron acceleration from active
regions on the sun, sending strong electron beams along open field lines into the heliosphere.

The case of type II radio bursts is still a subject of active research
\citep{KnockModel, schmidtCMEshocks, SolarWindGanse}, with models favouring electron drift acceleration on
CME shock fronts as the driver behind these events.

Common to both classes of radio bursts is the presence of a beamed electron
population in an otherwise quiescent plasma background, which leads to
excitation of electrostatic plasma wavemodes (Langmuir waves) through
Cerenkov-type instabilities \citep{pulupabale10}.
These electrostatic waves
then undergo three wave interaction processes which finally yield transverse
electromagnetic waves, which are observed as radio emissions: \begin{eqnarray}
	L &\rightarrow& L' + S \label{eq1} \\ L &\rightarrow& S + T(\omega_{pe})
	\label{lts}\\ L + L' &\rightarrow&T(2\omega_{pe}) \label{llt}
	\label{melrosecouplings} \end{eqnarray} Where $L$ denotes Langmuir-, $S$
	sound waves and $T$ transverse electromagnetic modes at the given frequency
	\citep{Melrose}.

Type II radio bursts are of particular merit for the fundamental study of
thee-wave interactions since they are the only naturally occurring three-wave
interaction process in which the emitting plasma can be probed in-situ by
satellite observations, and high-resolution radio observations of the resulting
waves are available.

\section{Kinematics of three-wave interaction}

Momentum and energy conservation impose certain limits on the waves
participating in three-wave interaction processes \citep{SpanierVainio09},
favouring certain wave combinations while entirely forbidding others.

In the case of beam-generated Langmuir-wave interaction, the Langmuir waves
(L) should primarily be generated parallel to the beam direction, whereas
electromagnetic waves are expected to be emitted in quasi-perpendicular
directions:

Assuming that \begin{eqnarray} k^L_\parallel &=& k_\parallel^T + k_\parallel^S
	\label{analytic1}\\ k^L_\perp &=& k_\perp^T + k_\perp^S\\ \omega^L &=&
	\omega^T + \omega^S \label{analytic2} \end{eqnarray} and inserting the
	dispersion relations for all three waves and neglecting $k_\parallel^T$ as
	well as $k_\perp^S$ due to angular momentum conservation
	\citep{SpanierVainio09}, we get \begin{equation} 3k_\parallel^2
		v^2_\mathrm{th} = c^2 k_\perp^2 + k_\parallel^2 v_s^2 + 2k_\parallel v_s
		\omega_{pe} + k_\perp k_\parallel v_s c, \label{analytic3} \end{equation}
		(with speed of sound $v_s$ and electron thermal speed $v_\mathrm{th}$)
		which results in an expected $k$-value for the electromagnetic emission of
		\begin{equation} k_\perp = -\frac{v_s}{2c} k_\parallel \pm \sqrt{3
			k_\parallel^2 \left( \frac{v_\mathrm{th}^2}{c^2} - \frac{v_s^2}{4 c^2}
			\right) - 2 \frac{v_s}{c^2}k_\parallel \omega_{pe}} \label{analytic4}
		\end{equation}

\section{The ACRONYM PiC-Code}

The Particle-in-Cell code ACRONYM (\textbf{A}dvanced \textbf{C}ode for
\textbf{R}elativistic \textbf{O}bjects, \textbf{N}ow with \textbf{Y}ee-Lattice
and \textbf{M}acroparticles) is being used in the numerical study of these
three wave interaction processes.  It is a fully relativistic MPI-parallelized
Particle-in-Cell code for 2d3v and 3d3v simulations.

The code is written in fully portable C++ and has successfully been used on
different architectures including x86-64, UltraSPARC and PowerPC. The code
efficiently scales from single core PCs up to supercomputers with more than
100000 cores.

Development started at the department of astronomy, University of W\"urzburg in
2007. Since then the code has seen numerous improvements in speed and accuracy.
It was used for simulation runs on our local clusters, the NEC Nehalem Cluster
at HLRS, the Linux Cluster and the Altix 4700 at LRZ, Louhi at CSC, Mare
Nostrum in Barcelona and Juropa and Jugene in J\"ulich.

All important numerical methods used in ACRONYM are accurate to second order.
Particles are represented by macroparticles with a triangular shaped cloud
(TSC) form factor. The particle motion is propagated using a (second order) Boris
push.

ACRONYM has been used for the simulation of a wide range of plasma conditions.
It was successfully used in the high-density regime of Laser Wakefield
Acceleration with length scales of a few nanometers, up to the case of
filamentation simulations where the length scales are hundreds of meters.
Large runs of the code regularly contain several billion particles.

\section{Numerical Representation of Waves}

A Particle-in-Cell code performs completely local calculations within the
spatially cell-partitioned simulation domain according to particle kinetics,
and as such carries no inherent concept of a wave. The wave behaviour is
completely emergent from the large-scale evolution of the simulated
Vlasov-Maxwell system.

\begin{figure}[htb] \begin{center}
	\includegraphics[width=6cm]{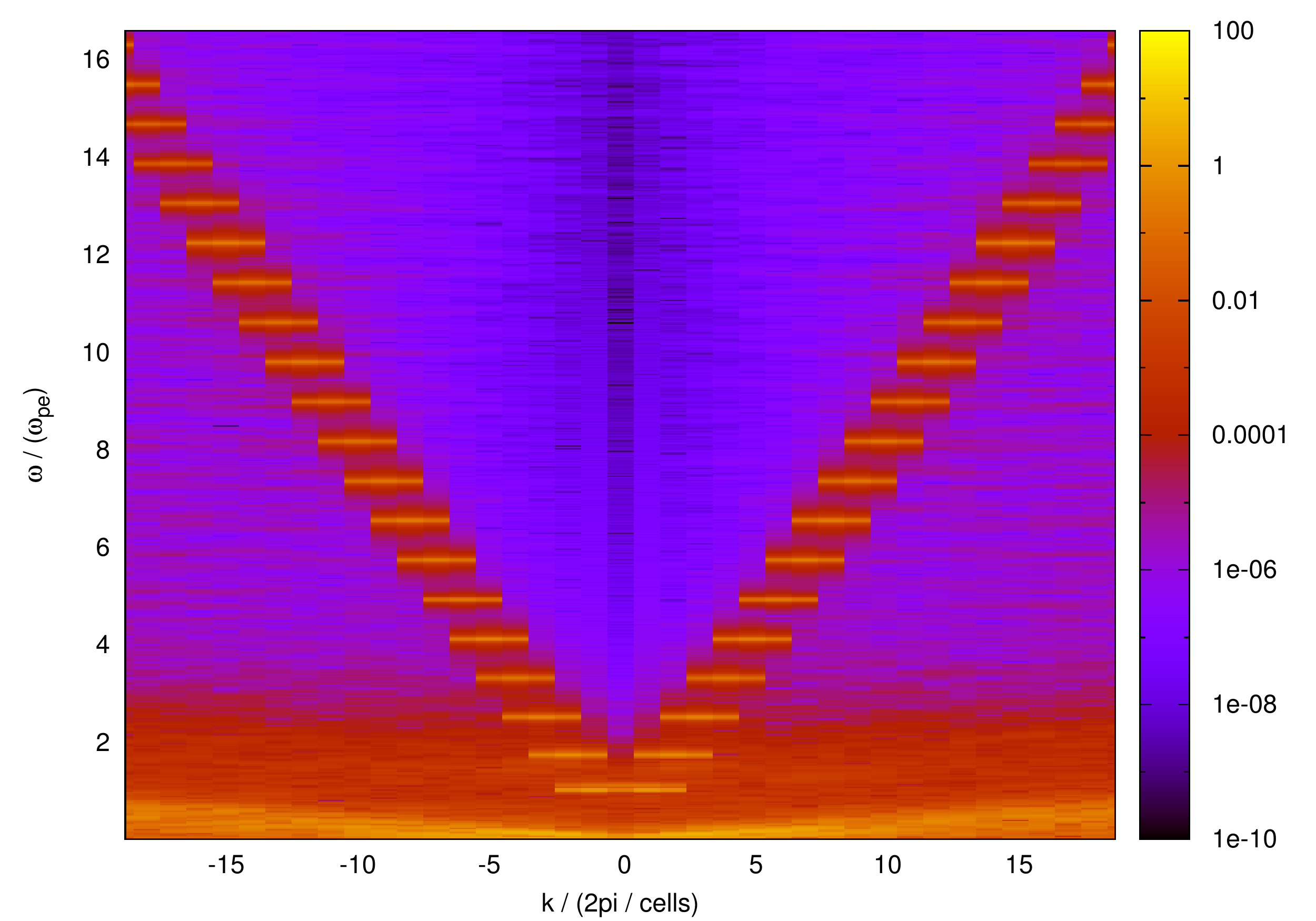}
	\includegraphics[width=6cm]{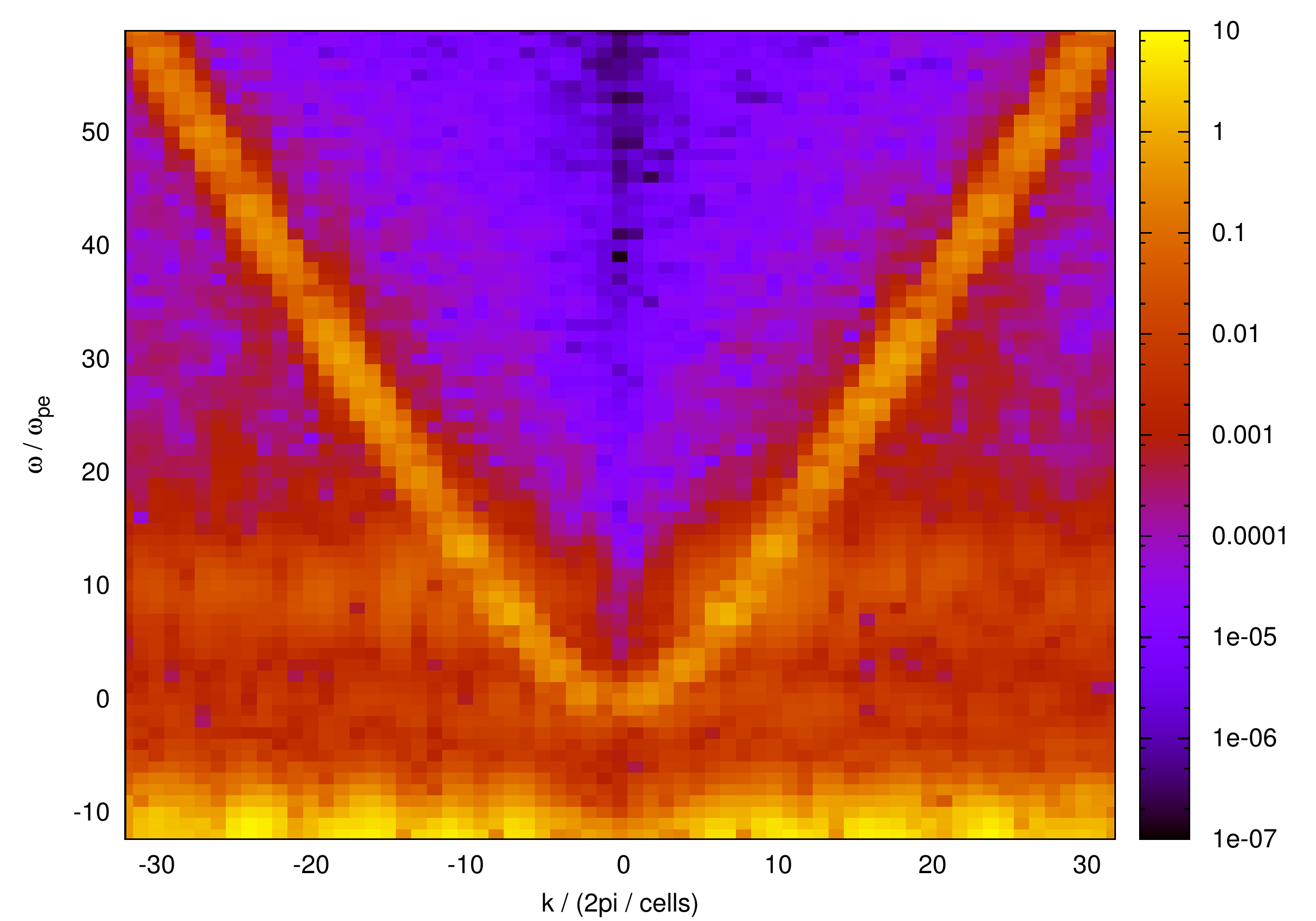} \end{center} \caption{Wave
	mode representation in a pic simulation with insufficient spatial resolution
	(left) compared to one with sufficient resolution. The physically continuous
	dispersion relation of an electromagnetic mode decomposes into discrete mode
	islands.} \label{fig:nicht_genug}
\end{figure}

In order to represent a wave with a certain wave-vector $\vec{k}$ within the
simulation, both sufficient resolution ($\Delta x < 2 \pi |k|^{-1}$) and
sufficient spatial extent ($\Delta x \cdot n_x > 2 \pi |k|^{-1}$) have to be
available. If this is not the case, the normally continuous wave dispersion
relation will transform into a set of discrete modes, which form multiples of
the inverse spatial length. In simulations with periodic boundaries, these
modes will also self-couple across the simulation domain, giving largely
unphysical intensities (this is shown in figure \ref{fig:nicht_genug}).

In the case of a type II radio emission foreshock region, these requirements
lead to resolutions of at least 4000 cells in longitudinal and transverse
direction to obtain sufficient wave representation. \cite{KarlickyVandas} used
a 1d3v code with a resolution of 10000 cells, which recreated all wavemodes
with $\vec{k}$ collinear with the beam direction. In order to represent also
waves in oblique and perpendicular directions, an extension to 2d3v has to be
performed.

Combining all these numerical constraints, the complete simulation has been
shown to require at least $8192 \times 4096$ cells, with a minimum of 200
particles per cell to limit numerical noise. The timescales of the wave
coupling processes observed herein demand simulation durations of at least
30000 timesteps, thus leading to a grand total of $201 \cdot 10^{12}$ particle
updates, each taking about $8\,\mathrm{\mu s}$ on a modern CPU. The
resulting computing time requirements of about 500000 CPU-hours per simulation
run (not counting I/O) basically demand large supercomputers to be feasible. If
the setup is to be expanded into a third spatial dimension, computation time
quickly becomes unmanageable.

Quantitative analysis of wavemodes in the simulation output is performed via
temporal and spatial Fourier transforms of the produced, spatially resolved
quantities, typically yielding $\omega - k$ dispersion plots. In these, either
intensity or phase of the corresponding waves can be displayed.

Due to the large simulation extents required for sufficient $k$-space
resolution, the amounts of data that will be used in these Fourier transforms
can be quite massive. In the example of our CME foreshock simulations, total
raw data sizes easily exceeded 1 Terabyte. While this can still be handled with a
fast Fourier transform in reasonable time, more advanced, less computationally
optimized statistical methods quickly become unusable.

Fourier transformation only yields information about the transfer of energy between
wavemodes in the simulation, without giving quantitative direct information
about the individual wave couplings. Wavelet Bicoherence
\citep{WaveletBicoherence} would be a suitable method for obtaining these
couplings, but can not be used in simulations of this size due to its
unfavourable computational scaling behaviour.

\section{Results}

\begin{figure}[thb]
	\begin{center}
		\includegraphics[width=8cm]{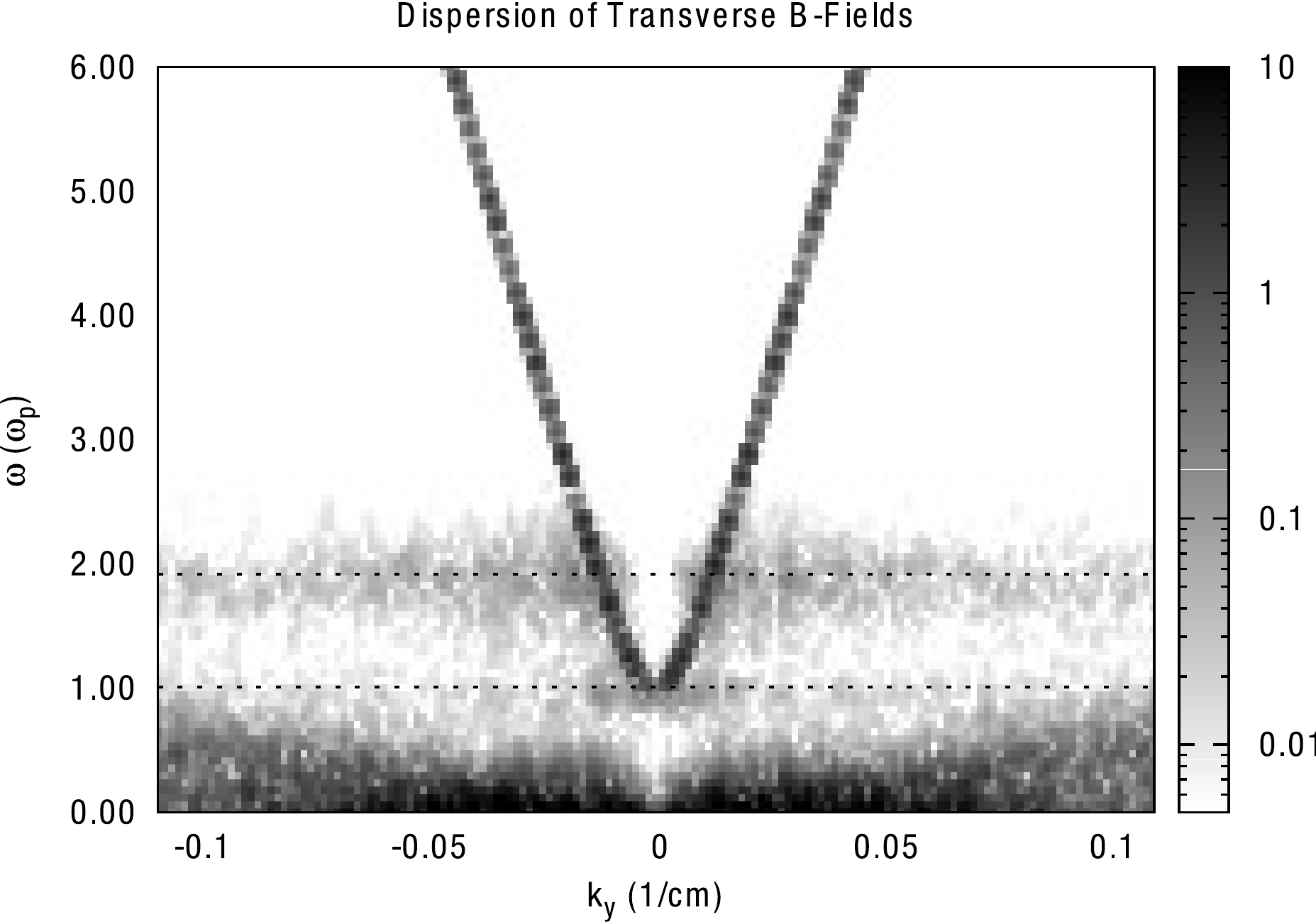}
	\end{center}
	\caption{Dispersion plot of a transverse magnetic field component from a
	foreshock pic simulation. The electromagnetic mode and nonlinear wave couplings at $\omega_p$ and $2 \omega_p$ are visible.}
	\label{fig:results}
\end{figure}

With 2d3v simulation extents of at least $8192\times4096$ cells, sufficient
resolution for the observation of Langmuir wave coupling to transverse
electromagnetic emissions were observed in the foreshock plasma simulations.
Figure \ref{fig:results} shows a dispersion diagram of a transverse magnetic
field component. The parabola-shaped electromagnetic mode predicted from linear
theory is clearly visible, as are low-frequency transverse waves (such as the
Alfv\'en wave), which are not properly resolved.

Additionally two bands of excitation are visible at $\omega = \omega_p$ and $2
 \omega_p$, which are consistent with the theoretically predicted three
wave interaction processes, and couple to the electromagnetic mode at the
resonance points.

Note that in this graph, only a small cutout of the complete dispersion plots
(in which the relevant physics is happening) is being displayed. Due to kinetic
resolution constraints, the much-finer resolved simulation leads to dispersion
plots which range to far higher spatial and temporal frequencies.

\section{Conclusions and further work}

The results presented here show a transfer in energy from beam-driven
Langmuir waves into transverse electromagnetic modes at both fundamental and
harmonic frequency, consistent with type II radio burst emission models. Since
the Fourier method employed for the analysis only yields information about the
intensity of the produced waves, quantitative information about the couplings
will need to rely on more advanced statistical methods. Effort is currently
ongoing to make bicoherence-based algorithms sufficiently scalable for dealing
with the data amounts produced in typical PiC simulation runs.

The simulation model will also be extended to compare setups with
counterstreaming electron beams (as described here) to single-beam setups,
in an attempt to confirm whether a single emission region or multiple beam
acceleration sites are necessary for type II radio bursts.

\acknowledgements 
The simulations in this research have been made possible through computing grants by
the Juelich Supercomputing Centre (JSC) and the CSC Computing Centre Helsinki. UG acknowledges support by the Elite Network of Bavaria.
This work has been supported by the European Framework Programme 7 Grant
Agreement SEPServer - 262773

\bibliography{ursg}

\end{document}